# WEAKLY NONLINEAR EVOLUTION OF TOPOLOGY OF LARGE-SCALE STRUCTURE[*]

Takahiko Matsubara

*Department of Physics, The University of Tokyo*
*Tokyo, 113, Japan*

and

*Department of Physics, Hiroshima University*
*Higashi-Hiroshima, 724, Japan*

## ABSTRACT

The gravitational evolution of the genus and other statistics of isodensity contours of the density field is derived analytically in a weakly nonlinear regime using second-order perturbation theory. The effect of final smoothing in perturbation theory on the statistics of isodensity contours is also evaluated. The resulting analytic expression for the genus is compared with $N$-body numerical simulations and exhibits a good agreement.

## 1. Introduction

There are many statistical tools to investigate the three-dimensional pattern of the galaxy distribution including two-point correlation function, power spectrum, cluster correlations, higher-order correlations, probability distribution function, etc. Among others, there is a class of statistics using a smoothed density field which cut the noisy property of galaxy distribution.

For example, the topology[1] of isodensity contours of those smoothed field is the popular one. The genus $G$, which is defined by $-1/2$ times the Euler characteristics of two-dimensional surfaces, divided by the total volume, can be a quantitative measure of the topology. The genus is a function of smoothing scales and the density threshold. The genus as a function of density threshold for a fixed smoothing scale is called the genus curve and is analyzed both in numerical simulations and in redshift surveys of galaxies by many people[2].

There are other statistics of isodensity contours, which include the 2D genus in two-dimensional slices of density field[3], $G_2$, the area of isodensity contours[4], $N_3$, the length of isodensity contours in two-dimensional slices[4], $N_2$, and the level crossing statistics[4], $N_1$.

---

[*]Talk presented at the 11th Potsdam Cosmology Workshop "Large Scale Structure in the Universe – Theoretical and Observational Aspects –", Potsdam, Germany, September 19 – 23, 1994

The analytical expression of these statistics are known for Gaussian random density field[3,4,5]. They are

$$G(\nu) = \frac{1}{4\pi^2}\left(\frac{\langle k^2 \rangle}{3}\right)^{3/2}(1-\nu^2)e^{-\nu^2/2}, \quad (1)$$

$$G_2(\nu) = \frac{1}{(2\pi)^{3/2}}\left(\frac{\langle k^2 \rangle}{3}\right)\nu e^{-\nu^2/2}. \quad (2)$$

$$2N_1(\nu) = \frac{4}{\pi}N_2(\nu) = N_3(\nu) = \frac{2}{\pi}\left(\frac{\langle k^2 \rangle}{3}\right)^{1/2}e^{-\nu^2/2}. \quad (3)$$

where $\nu$ is the difference between density threshold and mean density in units of standard deviation of density and

$$\langle k^2 \rangle = \frac{\int k^2 P(k) d^3k}{\int P(k) d^3k}, \quad (4)$$

with $P(k)$ being the power spectrum of the density fluctuation.

Previous analyses mainly compared the observed genus etc. with the random Gaussian prediction (1)–(3). With sufficiently large smoothing scales, this comparison could tell us if initial density fluctuation is random Gaussian or not. With the finite smoothing scale of cosmological interest, the effect of nonlinear gravitational evolution would be substantial. This nonlinear effect has been explored only by using $N$-body numerical simulations so far. In this paper, I present the analytic expression of statistics of isodensity contours extending the result for genus in the weakly nonlinear regime using a second-order perturbation theory[6].

## 2. Isodensity Statistics for Quasi-Gaussian Random Field

First, I introduce the seven quantities for a non-Gaussian random field $\delta(x,y,z)$ with zero mean as $(A_\mu) = \sigma^{-1}(\delta, \partial\delta/\partial x, \partial\delta/\partial y, \partial\delta/\partial z, \partial^2\delta/\partial x^2, \partial^2\delta/\partial y^2, \partial^2\delta/\partial x \partial y)$ where $\sigma \equiv \sqrt{\langle \delta^2 \rangle}$ is $rms$ of the field and the field is defined in Cartesian coordinates $x, y, z$. The statistics of isodensity contour $G, G_2, N_1, N_2, N_3$ of constant surfaces $\delta = \nu\sigma$ is given by[4,5]

$$G(\nu) = -\frac{1}{2}\left\langle \delta_D(A_1 - \nu)\delta_D(A_2)\delta_D(A_3)|A_4|(A_5 A_6 - A_7^2)\right\rangle, \quad (5)$$

$$G_2(\nu) = -\frac{1}{2}\left\langle \delta(A_1 - \nu)\delta_D(A_2)|A_3|A_5\right\rangle, \quad (6)$$

$$2N_1(\nu) = \frac{4}{\pi}N_2(\nu) = N_3(\nu) = 2\left\langle \delta(A_1 - \nu)|A_2|\right\rangle, \quad (7)$$

where $\delta_D$ is a Dirac's delta-function. This expression is valid for general homogeneous and isotropic non-Gaussian random fields.

In the following, I assume that $\langle A_{\mu_1} \cdots A_{\mu_N}\rangle_c \sim \mathcal{O}(\sigma^{N-2})$. This relation is a very definition of "weak non-Gaussianity" in this paper and is a result of perturbation theory[7]. Using the multi-dimensional version of the Edgeworth expansion,[†] and similar calculation developed by Matsubara[6], the above expression is expanded to the first order in $\sigma$ as follows:

$$G(\nu) = -\frac{1}{(2\pi)^2}\left(\frac{\langle k^2\rangle}{3}\right)^{3/2} e^{-\nu^2/2}$$
$$\times \left[H_2(\nu) + \sigma\left(\frac{S}{6}H_5(\nu) + \frac{3T}{2}H_3(\nu) + 3U H_1(\nu)\right) + \mathcal{O}(\sigma^2)\right], \quad (8)$$

$$G_2(\nu) = \frac{1}{(2\pi)^{3/2}}\frac{\langle k^2\rangle}{3} e^{-\nu^2/2}\left[H_1(\nu) + \sigma\left(\frac{S}{6}H_4(\nu) + T H_2(\nu) + U\right) + \mathcal{O}(\sigma^2)\right], (9)$$

$$2N_1(\nu) = \frac{4}{\pi}N_2(\nu) = N_3(\nu)$$
$$= \frac{2}{\pi}\left(\frac{\langle k^2\rangle}{3}\right)^{1/2} e^{-\nu^2/2}\left[1 + \sigma\left(\frac{S}{6}H_3(\nu) + \frac{T}{2}H_1(\nu)\right) + \mathcal{O}(\sigma^2)\right], \quad (10)$$

where $H_n(\nu) = (-)^n e^{\nu^2/2}(d/d\nu)^n e^{-\nu^2/2}$ are Hermite polynomials, and I have defined three quantities,

$$S = \frac{1}{\sigma^4}\langle\delta^3\rangle,$$
$$T = -\frac{1}{2\langle k^2\rangle\sigma^4}\langle\delta^2\nabla^2\delta\rangle, \quad (11)$$
$$U = -\frac{3}{4\langle k^2\rangle^2\sigma^4}\langle\nabla\delta\cdot\nabla\delta\nabla^2\delta\rangle.$$

The quantity $S$ is usually called "skewness". The first term in square brackets of Eqs. (8)–(10) corresponds to Gaussian contribution and the other terms correspond to non-Gaussian contribution.

## 3. Gravitational Evolution of the Genus Curve in Second Order Perturbation Theory

Gravitational nonlinear evolution give rise to $S$, $T$, $U$ even from the initial Gaussian random density fluctuation which has vanishing $S$, $T$, $U$. I use second order perturbation theory of the non-relativistic collisionless self-gravitating system in the fluid limit[9] to compute $S$, $T$, $U$ to lowest order in $\sigma$ in Einstein-de Sitter universe. The result including the effect of smoothing evolved field by Gaussian filter is given by Matsubara[6]. In fact, the observable curve is obtained by smoothing of density fluctuation at the final stages. Numerical values of $S$, $T$, $U$ for power-law initial spectra, CDM and HDM models are summarized in Table 1. In CDM and HDM models, I set $\Omega_0 = 1$, $H_0 = 50$ km s$^{-1}$ Mpc$^{-1}$ and smoothing length = 10 Mpc. The values of skewness $S$ in this table was first obtained by Juszkiewicz et al.[10]

---

[†]The usual one-dimensional Edgeworth expansion was recently applied to gravitational instability theory[8]

|   | unsmoothed | $n=1$ | $n=0$ | $n=-1$ | $n=-2$ | $n=-3$ | CDM | HDM |
|---|---|---|---|---|---|---|---|---|
| $S$ | 4.857 | 3.029 | 3.144 | 3.468 | 4.022 | 4.857 | 3.443 | 4.018 |
| $T$ | 3.905 | 2.020 | 2.096 | 2.312 | 2.681 | 3.238 | 2.335 | 2.970 |
| $U$ | 1.543 | 1.431 | 1.292 | 1.227 | 1.222 | 1.272 | 1.223 | 1.306 |

Table 1. The numerical values of $S$, $T$, $U$ for unsmoothed perturbation theory and smoothed perturbation theory for power-law spectra, $n = 1$ to $-3$, CDM and HDM models.

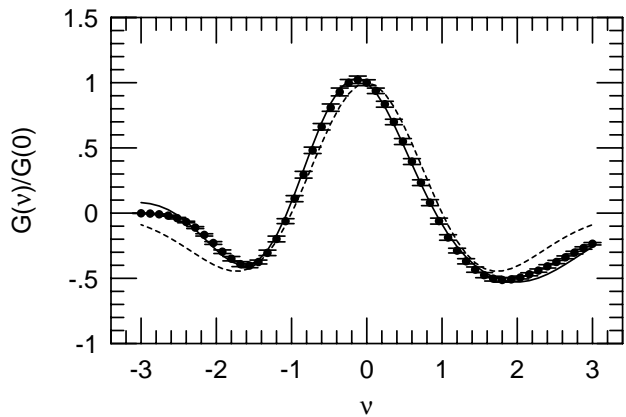

Fig. 1. Comparison of the analytic genus curve (a solid line) with that from an $N$-body simulation (symbols) for the Poisson model at $\sigma = 0.2$. A dashed line indicate the Gaussian prediction.

## 4. Comparisons with Numerical Simulations

Suto and I found that the analytic expression of the genus curve agrees well with the numerical $N$-body simulations for power-law initial fluctuations $P(k) \propto k^n$ and CDM models[11]. In Fig. 1 is plotted the normalized genus $G(\nu)/G(0)$ as an example of the comparison. The initial fluctuation is the Poisson model $P(k) =$ const. The *rms* of the fluctuation at the stage of comparison is $\sigma = 0.2$.

## 5. Discussion

The prominent feature of the results, Eqs. (8)–(10) is that weakly non-Gaussian correction introduces asymmetry to the symmetric or anti-symmetric curves. The pattern of the asymmetry is dependent on initial power spectra through smoothing effect of $S$, $T$, $U$. Thus, in principle, accurate observations on the statistics presented here can restrict the properties of initial fluctuation, such as Gaussianity, the shape of the spectrum, by the amplitude and the pattern of asymmetry of the curve. The projects as Sloan Digital Sky Survey (SDSS) will enable us to have a large amount of redshift data in near future and the analysis indicated in this paper will be important.

## 6. Acknowledgements

I am grateful to thank Y. Suto for useful discussions. I acknowledge the support of JSPS Fellowship. This research was supported in part by the Grants-in-Aid for Scientific research from Ministry of Education, Science and Culture of Japan (No. 0042).